# INFERENCE OF CHROMOSPHERIC MAGNETIC FIELDS IN A SUNSPOT DERIVED FROM SPECTROPOLARIMETRY OF Ca II 8542 Å

Ali G. A. Abdelkawy[1], Abdelrazek M. K. Shaltout[1], M. M. Beheary[1], T. A. Schad[2]

[1] Department of Astronomy and Meteorology, Faculty of Science, Al-Azhar University, Nasr City, Cairo, Egypt, Postal Code 11884.
[2] Department of Planetary Sciences, University of Arizona, Tucson, AZ 85721.

## Abstract

We analyze spectropolarimetric observations of the chromospheric Ca II 8542 Å line taken by the Interferometric Bidimensional Spectrometer (IBIS) at the Dunn Solar Telescope. The data were observed on 2012 January 29 for the NOAA active region 11408. Adopting the center-of-gravity (COG) approach we obtain the line-of-sight (LOS) field strength for the chromospheric IBIS data of Ca II 8542 Å line. The LOS strength of the magnetic field is determined in the target active region inside a field of view 45″ x 95″. The LOS field values were found to be increase up to 800 G inside the umbral region and decrease systematically toward the edges of a sunspot. Under the weak field approximation (WFA), the horizontal & vertical magnetic field components and azimuthal field vector are obtained.

## 1. Introduction

The solar active phenomena normally connected with the solar magnetic field. At the surface of the Sun, the magnetic field is focused in the areas called sunspots. These areas consist of dark regions at the center which is commonly called the umbra and a surrounding less dark region as called the penumbra. High spatial resolution observations of sunspots show that there are a lot of different morphological phenomena. Sunspots appear as a magnetic flux tube rise and intersects with the photosphere.

The IBIS is a two-dimensional spectropolarimeter that has been located for the polarimetry of the Sun (Cavallini 2006; Reardon & Cavallini 2008) established at the National Solar Observatory at the Dunn Solar Telescope (DST). It produces polarimetric data of the four Stokes IQUV profiles using for measurements of the magnetic field on the Sun. The spectropolarimetric data allow us a good possibility to derive all components of the field strength with high spatial resolution, as recently determined by Ichimoto & Shaltout (2012) and Shaltout & Ichimoto (2015).

In the current paper, we used the COG method are described by Rees & Semel (1979) to find the magnetic field strength from the observed Stokes I and V profiles without taking into account the filling factor of magnetic elements, within an accuracy of about 10%. It has been adopted in various investigations (Rees & Semel 1979; Cauzzi et al. 1993; Uitenbroek 2003; Balasubramaniam et al. 2004; Nagaraju et al. 2008; Kleint et al. 2009; Abdelkawy et al. 2016).

Also under the weak field approximation, the horizontal & vertical magnetic field components and azimuthal field vector of a sunspot are obtained.

## 2. Observations

The sunspot of NOAA 11408 was observed on 2012 January 29 with the IBIS. It was located at a heliocentric angle of ($\mu = \cos\theta = 0.8$). The IBIS recorded full Stokes profiles of Ca II line at 8542.1 Å (effective Landé factor g = 1.10) with a spatial sampling of 0″.0976. The observation started at 16:33 UT and ended at 17:04 UT, and its FOV was centered over the sunspot.

The spectral sampling of Ca II 8542 covered a range from 8540.35 Å to 8542.85 Å over the whole line profile. The IBIS image has a field of view (FOV) of 45″ x 95″. For more details on the observational data, it can be described in Abdelkawy et al. (2016).

Figure 1 shows the individual speckle reconstruction of the continuum intensity at the beginning of the sequence, Stokes I map for one wavelength position in the blue wing of the observed line and Stokes I map resulting in the Ca II 8542 line core.

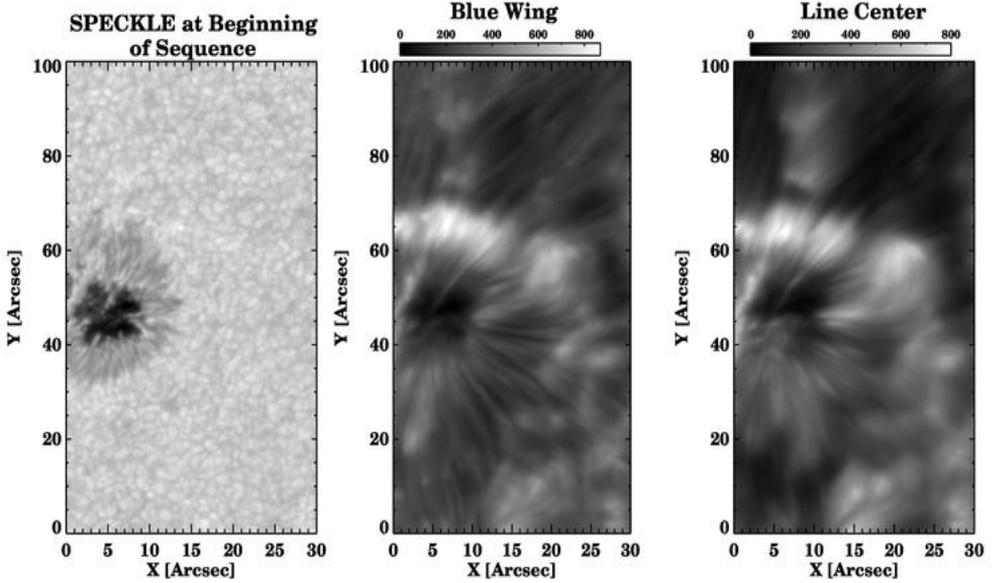

**Fig. 1.** The IBIS data of the Ca II 8542 of AR 11408 observed on 2012 January 29. In the left panel, an individual SPECKLE reconstruction of the continuum intensity, followed by the blue wing line intensity and line core in the Ca II 8542 line.

## 3. Center-of-gravity method

Several techniques exist for retrieving and studying information about the properties of the Solar atmosphere from the polarization state of spectral lines, which are commonly called spectral diagnostics. Apart from the commonly used of spectral diagnostics, Stokes I and V profiles are adopted in both sunspots and plage using the center-of gravity method for inferring chromospheric magnetic fields from high-resolution spectropolarimetric observations, as has been described in Abdelkawy et al. (2016).

The COG method adopted the I+V and I-V to derive the line-of-sight magnetic field strength ($B_{LOS}$) in each point of the field-of-view (FOV), where the wavelength difference of the COG of I+V and I-V is directly related to $B_{LOS}$ of magnetic field strength. Since the COG method can be used to the widely varying shapes of the chromospheric Ca II 8542 V-profiles, where the full FOV is completely used. The LOS magnetic field strength is calculated from Abdelkawy et al. (2016):

$$B_{LOS} = \frac{(\lambda_+ - \lambda_-)/2}{4.67 \times 10^{-13} \lambda_0^2 \, g_L} \qquad (1)$$

where ($\lambda_\pm$) are the positive and negative circularly polarized wavelengths, $g_L$ is the effective Landé factor and ($\lambda_0$) is the central wavelength of the line in (angstrom).

The integration is over the spectral range of a given spectral line. Figure 2 presents the COG map of Ca II 8542 Å line and average field value of wide concentric rings around the sunspot center and intensity as functions of the radial distance from the center of a sunspot, derived by the COG method, respectively. The chromospheric field strength ($B_{LOS}$) is systematically increase at the umbral region until a point close to the outer boundary of the penumbra where the relation changes toward the edges of sunspot. This feature of the field strength a function of the radial distance is in agreement with that of Kleint et al. (2009) result.

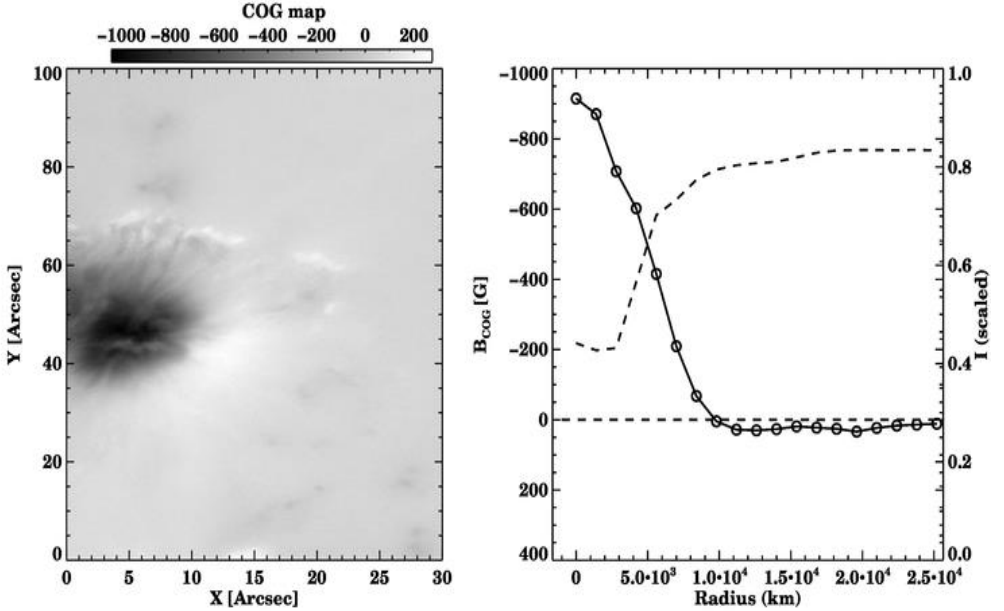

**Fig. 2.** The COG map of Ca II 8542 Å line of a sunspot AR 11408. (left panel) and chromospheric magnetic field derived by the COG method (solid line) and intensity (dotted lines) as functions of the radial distance from the center of a sunspot (right panel).

## 4. Magnetic field components

When the Doppler width is larger than the Zeeman splitting the spectral line is in the weak field regime. In this case the longitudinal component of $B_{LOS}$ is constant with depth. Under the WFA, the Stokes V profile can be expressed as functions of the first or second derivative of the intensity profile (I) and the magnetic field. When using the full-Stokes observed profiles of the Ca II 8542 chromospheric line the field may vary more or less over the pixels and this is mainly due to technical limitations of the instrumentation. We derive the vertical (line of sight) component of the magnetic field ($B_{\parallel}$) (Hammar 2014) as defined by:

$$B_{\parallel} = \frac{-\sum_i V_i (\frac{\partial I}{\partial \lambda})_i}{C_1 f \sum_i (\frac{\partial I}{\partial \lambda})_i^2} \qquad (2)$$

where $V_i$ is the stokes V, I is the intensity, $\lambda$ is the sets of wavelength, $f$ is the filling factor and $C_1 = 4.6686 \times 10^{-13} \times \lambda_0^2 \times \bar{g}$, where $\lambda_0$ is the central wavelength, $\bar{g}$ is the effective landé factor, and it assumed $\bar{g} = 1.10$ for Ca II 8542 line. As noticed from figure 3 (right panel) that we find a similar trend of the magnetic field as obtained from the COG method. This supports the result of Kleint et al. (2009).

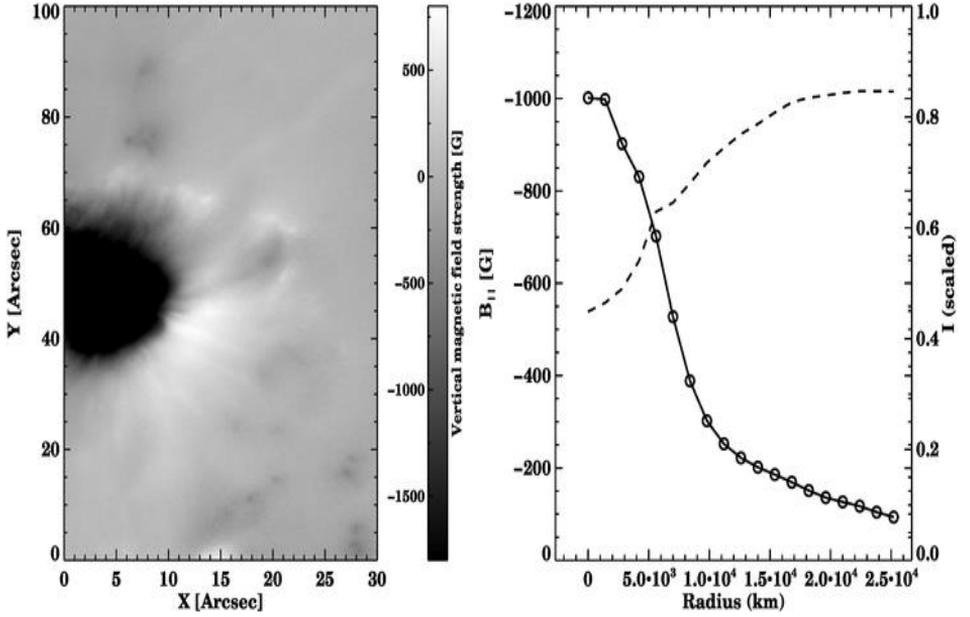

**Fig. 3.** Inferred line-of-sight vertical component of magnetic field strength image (left panel) and an average magnetic field values for a part of ellipse around the sunspot center (solid line) and the intensity (dotted line) as a function of the radial distance (right panel) of IBIS of AR 11408 observed on 2012 January 29.

We derived the horizontal (line of sight) component ($B_h$) of the magnetic field by this expression:

$$B_h = \left(\frac{\sum_i L_i \left|\frac{\partial^2 I}{\partial \lambda^2}\right|_i}{C_2 f \sum_i \left|\frac{\partial^2 I}{\partial \lambda^2}\right|_i^2}\right)^{1/2} \quad (3)$$

where $L_i$ is the total linear polarization ($L = \sqrt{Q^2 + U^2}$), where U, Q are the stokes U and stokes Q and the constant $C_2 = 5.4490 \times 10^{-26} \times \lambda_0^4 \times \overline{G}$, where ($\overline{G} = 1.22$), as described by Hammar (2014). We find a positive trend of the magnetic field as functions of the radial distance from the sunspot center. The horizontal field component is increase at the umbral region and it is systematically decrease at the edge of the sunspot where it is up to 1000 G. Our results of the vertical magnetic field component ($B_{\parallel}$) are up to 1600 G where a filling factor of unity was assumed, while for the horizontal component ($B_h$) ranges up to 3600 G. Our results are in agreement with that of Hammar (2014).

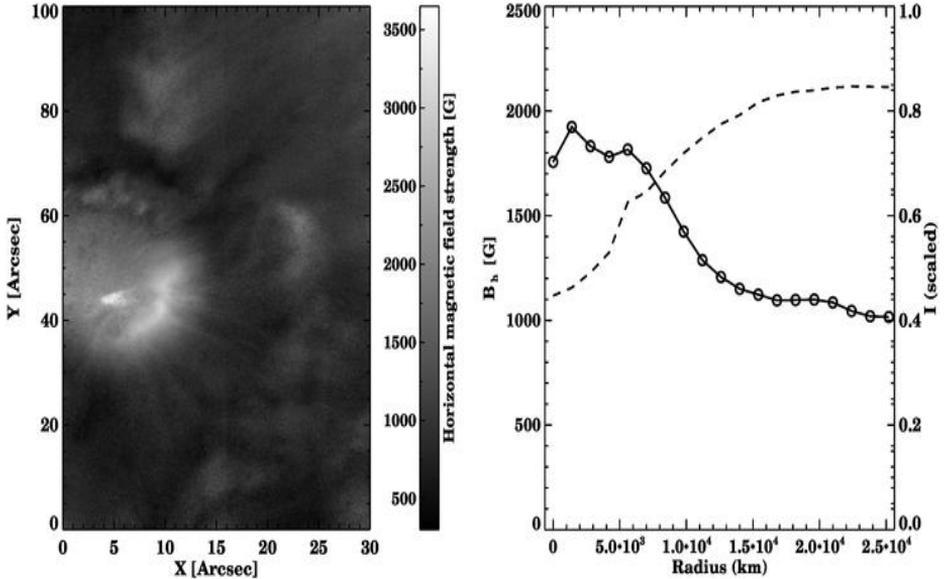

**Fig. 4**. The inferred the line-of-sight horizontal component of magnetic field (left panel) and an average magnetic field values (solid line) and intensity (dotted line) for a part of ellipse around the sunspot center as a function of the radial distance (right panel) of IBIS data.

We make a comparison of the line-of-sight magnetic field ($B_{COG}$) derived from the COG method and vertical component ($B_{\parallel}$) of magnetic field strength under the weak field regime of IBIS observations of the sunspot region, as shown in figure 5.

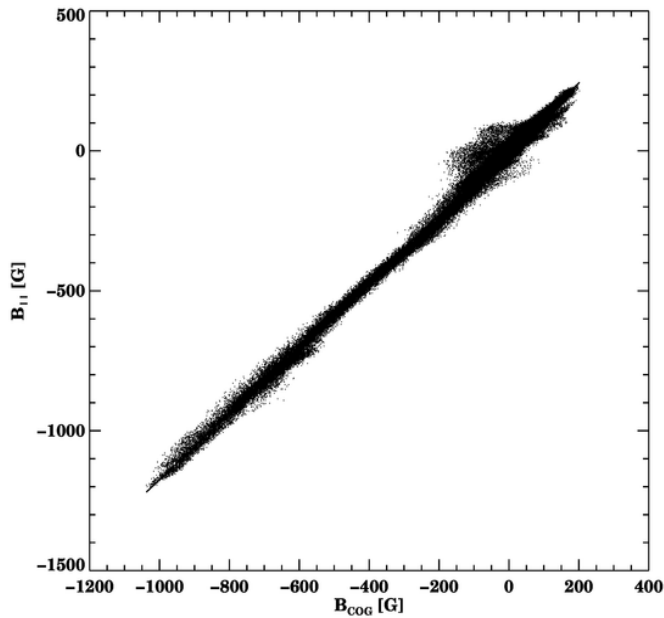

**Fig. 5.** A comparison between the line-of-sight magnetic field ($B_{COG}$) inferred from the COG method against the vertical component of magnetic field strength ($B_{\parallel}$) calculated using the weak field approximation.

## 5. Azimuth field vector

The azimuthal field vector ($\chi$) can be calculated by least-squares minimization, as suggested by Hammar (2014), where we used this formula:

$$\tan 2\chi = \frac{\sum_i U_i (\frac{\partial^2 I}{\partial \lambda^2})_i}{\sum_i Q_i (\frac{\partial^2 I}{\partial \lambda^2})_i} \qquad (4)$$

where the parameters U, Q are the stokes U and Q, I is the intensity of the line and $\lambda$ is the wavelength.

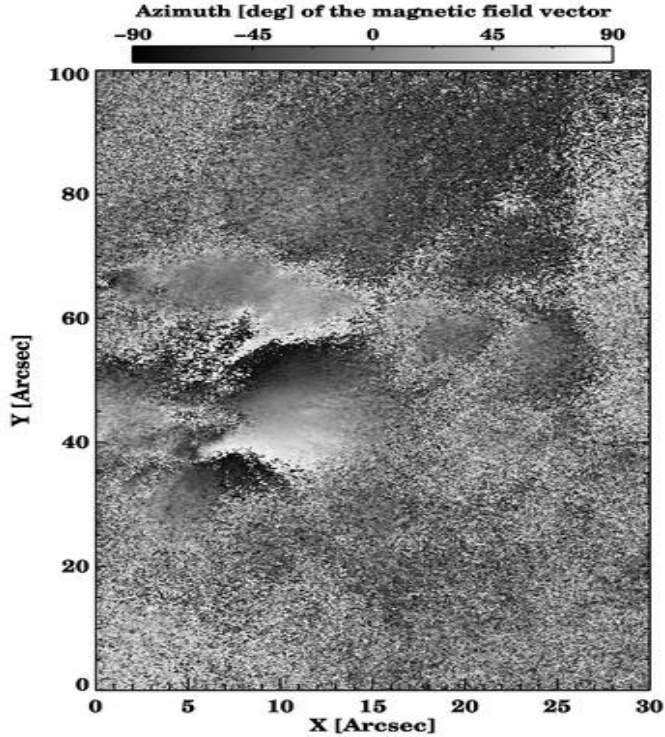

**Fig. 6.** The deduced azimuthal ($\chi$) field vector of the IBIS data of a sunspot.

Figure 6 shows the azimuthal field vector of a sunspot. Our results are in the range between of -90-to-90 deg., which is consistent with that of Schad et al. (2012).

## 6. Conclusions

In this paper, we inferred the LOS chromospheric magnetic field strength ($B_{COG}$) of IBIS observations of the observed sunspot region (AR 11408) on 2012 January 29, using the COG method (Rees & Semel 1979), our result is from -1 kG to 0.25 kG, this analysis was obtained without taking into account the filling factor of magnetic elements.

Under the weak field regime, the horizontal ($B_h$) & vertical magnetic ($B_{||}$) field components and azimuthal field vector are presented, where images of these parameters are analyzed with the radial distance from the sunspot center except for the azimuth field vector. The result of the vertical magnetic component ($B_{||}$) is up to 3600 G with a filling factor of unity, while the result of the horizontal component ranges up to 3600 G and it agrees with that of Hammar (2014). Finally, the azimuthal field vector of a sunspot is between the range -90-to-90 deg., which it is consistent with that of Schad et al. (2012). The observed Stokes profiles are induced by the Zeeman effect in transitions where the inversion analysis is a future project.

**Acknowledgments.** The National Solar Observatory in (U.S.A) is operated by the Association of Universities for Research in Astronomy, Inc., under cooperative agreement with the National Science Foundation. IBIS is a project of INAF/OAA with additional contributions from University of Florence and Rome and NSO. Gianna Cauzzi of National